\documentclass[twocolumn,showpacs,preprintnumbers,amsmath,amssymb,prl]{revtex4}

\usepackage{graphicx}
\usepackage{dcolumn}
\usepackage{bm}

\newcommand{\ket}{\rvert\psi_\alpha\rangle}
\newcommand{\ua}{_\alpha}
\newcommand{\oa}{^\alpha}
\newcommand{\bra}{\langle\psi_\alpha\lvert}
\newcommand{\conv}{\otimes}
\newcommand{\dr}{d^2\mathbf{r}}

\begin{document}


\title{Nonlinear management of the angular momentum of soliton clusters}

\author{Andrea Fratalocchi}
 \email{frataloc@uniroma3.it}

\author{Armando Piccardi, Marco Peccianti}

\author{Gaetano Assanto}
 \email{assanto@uniroma3.it}
\affiliation{
Nonlinear Optics and OptoElectronics Labs (NooEL),\\
INFN and CNISM, University ROMA TRE,
Via della Vasca Navale 84, 00146, Rome, Italy
}

\date{\today}

\begin{abstract}
We demonstrate an original approach to acquire nonlinear control over the angular momentum of a cluster of solitary waves. Our model, derived from a general description of nonlinear energy propagation in dispersive media, shows that the cluster angular momentum can be adjusted by acting on the global energy input into the system. The phenomenon is experimentally verified in liquid crystals by observing power-dependent rotation of a two-soliton cluster.
\end{abstract}


\maketitle

\paragraph{Introduction. ---}
\label{intro}
Angular momentum (AM) is a fundamental quantity, the importance of which has been highlighted in almost all areas of physical sciences. Evidence of its role is found at the inception of the universe: although the distribution of  galaxies, stars and planets is still a puzzle in astrophysics, it appears that an initial angular momentum in the early universe prevented cluster-sized clouds from collapsing into a series of black holes, i.e. with no  planets to support life \cite{PBTMOCD,am_bf_kazana}. In both classical and quantum mechanics, angular momentum is at the basis of rotational dynamics; hence, AM governs the behavior of important processes such those arising from fluid motion (e.g. initiation of cyclones \cite{am_bf_mcewaa} and fluctuations in the length of a day \cite{bf_am_hide}), statistical complexes of rotating molecules \cite{bf_am_oord}, quantum optics \cite{Torner_PRA_03} and quantized particle ensembles. In the latter area, research has been mostly conducted in two major directions of investigation:\\
i) the revolution of trapped particles by their interaction with classical fields carrying angular momentum, fostering  nanotechnological applications such as rotors or more complex machines powered by light \cite{am_mmc_tabosa};\\
ii) the preparation of energy packets in well defined AM states, with implications to both fundamental physics \cite{bf_amc_minns} and quantum information systems \cite{bf_am_maira}.\\
However, despite the importance of AM in physics and the vast literature on the subject, neither methods to nonlinearly control AM have been proposed, nor thorough studies have been carried out on the effect of nonlinearity on the angular momentum of a specific system.  
Conversely, great attention has been devoted in the past few years to solitons and solitary waves. Such waves are ubiquitous and rely on the balance between wave-packet dispersion (spreading) and nonlinearity \cite{LANW}. Following the pioneering numerical experiments by Fermi, Pasta and Ulam \cite{sol_bf_fermia}, the universal concept of soliton has acquired importance in several physical sciences, including biology \cite{SICMP}, hydrodynamics \cite{LANW}, plasma physics \cite{sol_pl_Zabus}, ultracold atoms \cite{sol_bec_Burge}, optics \cite{OSFFTPC}, gravitation \cite{GS} and beyond \cite{Hoffmann}. By virtue of their robustness, solitons have potentials in applications  --- from optical telecommunications to atomic interferometry in Bose-Einstein condensates (BEC)--- and are the subject of vigorous theoretical studies \cite{NWAS}. Recently, the interactions between two-dimensional deterministic soliton clusters have attracted interest, encompassing a wealth of dynamics ranging from wave filamentation \cite{fl_bf_bigela}, to spinning \cite{Mih_2002} to spiraling \cite{sp_bf_belica,to_ss_rotsca,bf_sc_desyaa}. Up to date, however, studies focused on either specific nonlinear models or particular input waveforms \cite{fl_bf_bigela, sp_bf_belica,to_ss_rotsca,bf_sc_desyaa}, or on 1D stochastic dynamics \cite{bf_cpxl_conti, pecciante_OL}; none of them discussed the role of the excitation on the dynamics of a two-dimensional multi-soliton ensemble.\\
In this Letter, by employing a rather universal model for the theory and a nonlocal dielectric for the experiments, we investigate the behavior of a cluster of (2+1)D optical solitons, demonstrating a nonlinear approach to controlling its angular momentum. For the sake of simplicity but with no prejudice on its general validity, we develop the analysis in the simplest case of a two-soliton cluster, starting from first principles and employing the language of symmetries \cite{AOLGTDE}. We demonstrate that \emph{the angular momentum of a soliton cluster exhibits a linear dependence on the nonlinear response; hence it can be precisely managed by varying the input power. The AM can therefore be evaluated from the global revolution of the cluster which, for a fixed propagation distance, evolves linearly with excitation}. We check the theoretical results by a series of experiments in nematic liquid crystals (NLC), a nonlinear nonlocal medium known to support stable (2+1)D solitons in voltage-tunable configurations. \cite{nlc_st_pecci,non_BangS, non_NLC_ContT}. 
\paragraph{Theory. ---}
We build a general model of nonlinear wave propagation, stemming from conservation laws and variational symmetries (as there is a one-to-one correspondence between them \cite{AOLGTDE}) and considering an isolated medium with translational and rotational invariance. The medium exhibits an optical  response nonlinear with the wave intensity; hence it enforces the conservation laws of momentum ($\mathbf{M}$), angular momentum ($\mathbf{A}$), Hamiltonian ($\mathcal{H}$) and energy flux ($\mathcal{W}$), respectively. Such a nonlinear model is derived by defining a suitable action integral $\mathcal{I}=\int \dr dz \mathcal{L}$, the Lagrangian of which supports the variational symmetries originated by the basis of Lie generators $\{\mathbf{v}_1=\partial/\partial x,\mathbf{v}_2=\partial/\partial y,\mathbf{v}_3=\partial/\partial z,\mathbf{v}_4=x\partial/\partial y- y\partial/\partial x,\mathbf{v}_5=i\psi_\alpha\partial/\partial\psi^\alpha+\mathrm{c.c.} \}$ (we adopt Einstein's  summation over repeated indices), being $\mathbf{r}=[x,y]$ and $z$ dimensionless coordinates and $\psi_\alpha$ the dimensionless wave-function of the $\alpha-$th wave-packet $(\alpha\in [1,2]$ in this work). A general form for the Lagrangian $\mathcal{L}$ is as follows:
\begin{align}
  \label{alag0}
\mathcal{L}=\frac{1}{2}\bigg(i\psi_\alpha^*\frac{\partial\psi^\alpha}{\partial z}&+\mathrm{c.c}\bigg)-\frac{1}{2}
\nonumber\\
&\times\nabla\psi\ua^*\nabla\psi\oa+\frac{1}{2}\lvert\psi^\alpha\rvert^2\Theta_{\alpha\beta}\conv\lvert\psi^\beta\rvert^2
\end{align}
with $\alpha,\beta\in [1,2]$, $\nabla=[\partial/\partial x,\partial/\partial y]$, $\Theta_{\alpha\beta}$ the Hermitian tensor and $\conv$ a convolution operator defined by $f\conv g=\iint d^2\mathbf{r'}f(\mathbf{r'}-\mathbf{r})g(\mathbf{r'})$. The general character of the Lagrangian (\ref{alag0}) can be proven from the Euler-Lagrange equations of motion, which read:
\begin{align}
\label{ael0}
  i\frac{\partial\psi_\alpha}{\partial z}+\frac{1}{2}\nabla^2\psi_\alpha+\psi_\alpha\Theta_{\alpha\beta}\conv\lvert\psi^\beta\rvert^2=0
\end{align}
The linear portion of Eq. (\ref{ael0}) is Schr\"odinger like and describes wave-packet propagation in the presence of dispersion \cite{WII}; the nonlinear portion ($\Theta_{\alpha\beta}\conv\lvert\psi^\beta\rvert^2$) can model both local ---i.e. for $\Theta_{\alpha\beta}(\mathbf{r})=c_{\alpha\beta}\delta (\mathbf{r})$--- \cite{OSFFTPC} and nonlocal \cite{non_NLC_ContT,sol_bec_Burge,pr_nl_christ} responses. By construction, the Lagrangian (\ref{alag0}) admits  the symmetries generated by $\mathbf{v}_i$ ($i=[1,5]$); hence Eq. (\ref{ael0}) possesses the following integrals of motion:
\begin{align}
  \label{aim0}
&\mathbf{M}=\iint\dr(i\psi_\alpha^*\nabla\psi^\alpha+\mathrm{c.c.})\\
&\mathcal{H}=\frac{1}{2}\iint\dr(\lvert\psi^\alpha\rvert^2\Theta_{\alpha\beta}\conv\lvert\psi^\beta\rvert^2-\nabla\psi\ua^*\nabla\psi\oa)\\
&\mathbf{A}=\iint\dr[\mathbf{r}\times(i\psi_\alpha^*\nabla\psi^\alpha+\mathrm{c.c.})]\\
\label{aim1}
&\mathcal{W}=\iint\dr\lvert\psi_\alpha\rvert^2
\end{align}

\begin{figure}
\includegraphics[width=8.5cm]{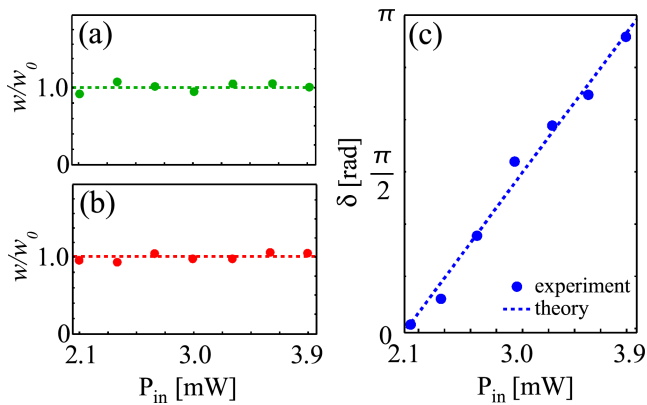}
\caption{\label{spir1}
(Color online). (a-b) Measured output spot-size $w$ of the two solitons normalized to $w_0=w(2.1mW)$ and (c) revolution angle $\delta$ versus input power $P_\mathrm{in}$.  
}
\end{figure}

The conserved quantities (\ref{aim0})-(\ref{aim1}) allow to generalize the Ehrenfest theorem which, with reference to Eq. (\ref{ael0}), takes the form:
\begin{align}
  \label{aer0}
&\frac{\partial}{\partial z}\bra\mathbf{r}\ket=\bra\mathbf{p}\ket
\nonumber\\
&\frac{\partial}{\partial z}\bra\mathbf{p}\ket=\bra\nabla\Theta\ua\ket 
\end{align}
being $\Theta\ua=\Theta_{\alpha\beta}\conv\lvert\psi^\beta\rvert^2$, $\mathbf{p}=-i\nabla$ the momentum operator and $\langle f_\alpha\lvert g_\alpha\rangle=\iint\dr f_\alpha^*g_\alpha$ the inner product defining the space metric. The set (\ref{aer0}) is a generalization of models either derived in one dimension \cite{bf_cpxl_conti} or obtained through linear expansions \cite{bf_sc_garcia,sp_bf_belica,bf_sc_desyaa}. Out of the whole spectrum of solutions to Eq. (\ref{ael0}), we are interested in the evolution of a nonlinear cluster, the solitary ``particles'' (solitons) of which are found as invariant solutions of Eq. (\ref{ael0}) (written for $\alpha=\beta$) with respect to the global symmetry group generated by $\mathbf{v}=\mathbf{v}_3+\mathbf{v}_4-x_\alpha\mathbf{v}_1+y_\alpha\mathbf{v}_2+\mu\mathbf{v}_5$. Then the method of characteristics \cite{AOLGTDE} yields the functional form of each solitary wave:
\begin{equation}
  \label{asol0}
 \psi_\alpha(\mathbf{r};z)=\phi (\lvert\mathbf{r}-\mathbf{r}_\alpha\rvert)\exp(i\mu z) 
\end{equation}
being $[x_\alpha,y_\alpha]=\mathbf{r}_\alpha=\bra\mathbf{r}\ket/\iint\dr\lvert\psi_\alpha\rvert^2$ the soliton ''center of mass''. The term ``particle'' is justified as long as the solitons do not collide: the nonintegrable nature of Eq. (\ref{ael0}), in fact, does not guarantee that after collision(s) the soliton ensemble evolves iso-spectrally \cite{bf_solg_ela}, i.e. with a constant number of wavepackets. However, if solitons simply overlap, the effects of the interaction terms ($\Theta_{\alpha\beta}\conv\lvert\psi_\beta\rvert^2$ for $\alpha\neq\beta$) can be treated adiabatically, i.e. by assuming that the latter do not affect the functional form of the solitons but only their phase $\mu$ or their center of mass $\mathbf{r}\ua$. Such hypothesis allows us to cast Eq. (\ref{aer0}) in a potential form. By substituting Eq. (\ref{asol0}) into (\ref{aer0}), after some straightforward algebra, we obtain:
\begin{align}
\label{apot0}
  m\frac{\partial^2\mathbf{q}}{\partial z^2}+\nabla_{q}U(\mathbf{q})=0
\end{align}
being $m=\iint\dr\lvert\psi_\alpha\rvert^2$ the soliton ``mass'', $\mathbf{q}=\mathbf{r}_1-\mathbf{r}_2=[q_x,q_y]$, $\nabla_{q}=[\partial/\partial q_x,\partial/\partial q_y]$, $U(\mathbf{q})=-[U_{12}(\mathbf{q})+U_{21}(\mathbf{q})]$ and:
\begin{align}
  \label{apot1}
&U_{\alpha\beta}=\iint\dr'\phi(\lvert\mathbf{r'}+\mathbf{q}\rvert)^2\Theta_{\alpha}(\lvert\mathbf{r'}\rvert), &\alpha\neq\beta
\end{align}
Equation (\ref{apot0}) describes the classical motion of a mass point (of mass $m$) subject to the nonlinear potential defined by (\ref{apot1}). It is worth to remark that we do not rely on a linear expansion; hence, Eq. (\ref{apot0}) holds true also when the excitation varies. The Lagrangian density of (\ref{apot0}) is $\mathcal{L}=\mathbf{\dot{q}}\mathbf{\dot{q}}/2-U(\mathbf{q})$ and admits the variational symmetries generated by the basis $\{\mathbf{v}_1=\partial/\partial z,\mathbf{v}_2=q_x\partial/\partial q_y-q_y\partial/\partial q_x\}$, the latter originating the conservation laws of energy $\mathcal{E}$ and angular momentum $\mathbf{L}$, respectively:
\begin{align}
  \label{acl1}
&\mathcal{E}=m\frac{\mathbf{\dot{q}}\mathbf{\dot{q}}}{2}+U(\mathbf{q})
\\
 \label{acl2}
&\mathbf{L}=m\mathbf{q}\times\mathbf{\dot{q}}
\end{align}
\begin{figure}
\includegraphics[width=8.5 cm]{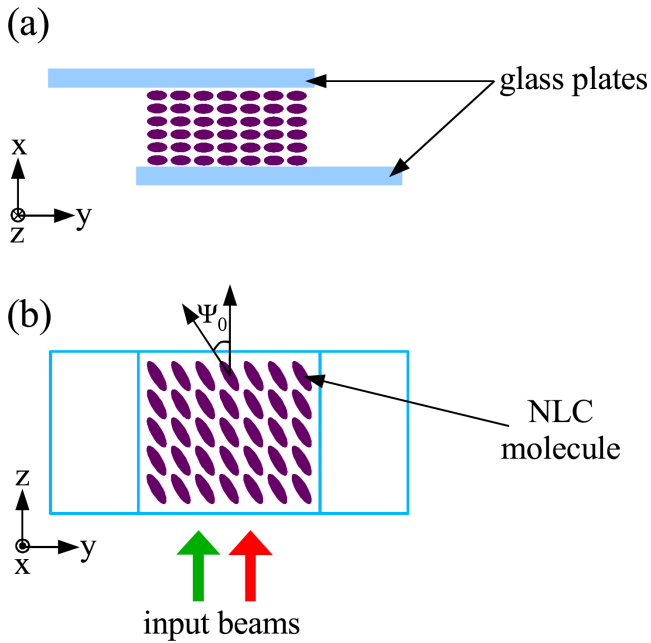}
\caption{\label{scell}
(Color online). Sketch of the planar cell with nematic liquid crystals: (a) front view and (b) top view with an indication of the molecular orientation (optic axis) in the plane $(y,z)$.
}
\end{figure}
Noticeably, equation (\ref{acl2}) states that:\\
i) the angular momentum is conserved, hence it is controlled by the system input;\\
ii) the angular momentum depends linearly on the soliton ``mass'' $m$, hence, on the nonlinear potential $U$.\\
As a consequence, a change in the soliton \textit{global} power $\iint\dr\lvert\psi\rvert^2=\iint\dr(\lvert\psi_1\rvert^2+\lvert\psi_2\rvert^2)$ results in a linear variation of AM, provided the adiabatic condition [i.e. Eqs. (\ref{asol0})-(\ref{acl1})] holds. To measure the AM observable, hereby we suggest an original approach exploiting soliton  \textit{spiraling}. Such dynamics occurs when the initial momentum $\mathbf{\dot{q}}(0)=\mathbf{\dot{r}}_1(0)-\mathbf{\dot{r}}_2(0)$ balances the attractive force provided by the bound potential $U(\mathbf{q})$ \cite{sp_bf_belica}, resulting in a rigid rotation of the (two) soliton ensemble with constant separation $d$ and invariant angular velocity $\omega=\mathcal{L}/\mathcal{M}=\partial \delta/\partial z$, being $\delta$ the angle spanned by $\mathbf{q}$, $\mathcal{M}$ the ``momentum of inertia'' of the classical system (\ref{apot0}) \cite{sp_bf_belica,bf_sc_desyaa}. Following a straightforward integration of $\mathcal{L}/\mathcal{M}=\partial \delta/\partial z$, a measure of soliton cluster AM is given by the revolution angle $\delta$ which, for a given $z$, varies linearly with the angular momentum $\mathcal{L}$. As a result, according that the adiabatic condition holds, we expect that the revolution angle $\delta$ of the soliton ensemble evolves linearly with the soliton input power $P_\mathrm{in}=\iint\dr\lvert\psi\rvert^2$.
\paragraph{Experiments. ---}
\begin{figure}
\includegraphics[width=8 cm]{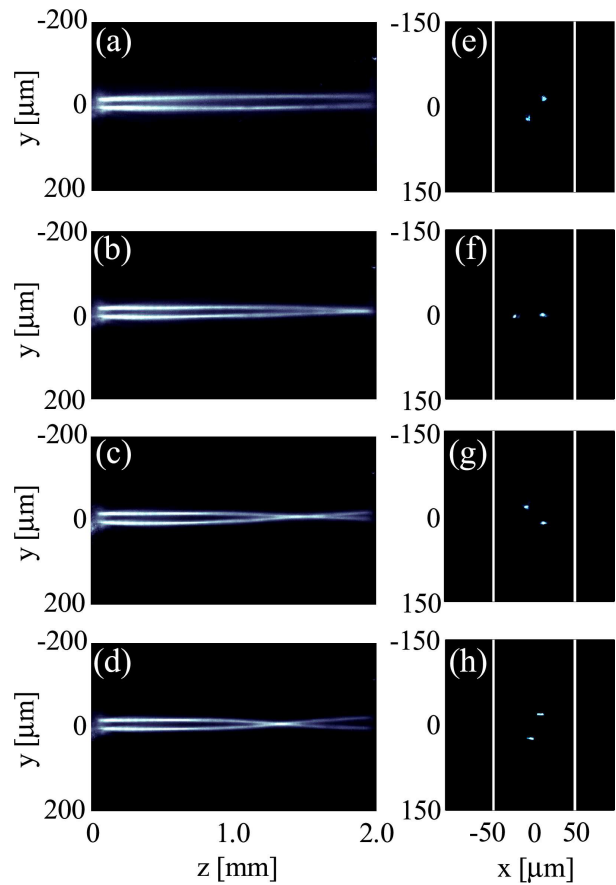}
\caption{\label{spir0}
(Color online). Summary of the experimental results: (a-d) evolution in the plane ($y,z$) and (e-h) output intensity profiles of the two-soliton cluster for increasing input powers $P_\mathrm{in}\in[2.1,2.7,3.3,3.9]mW$. The borders of the NLC cell are indicated with a solid line in (e-h).
}
\end{figure}
In the experiments we used a planar glass cell with a $100\mu$m thick layer of planarly anchored E7 liquid crystals (Fig. \ref{scell}). This configuration is similar to that previously employed for the investigation of accessible solitons in highly nonlocal media \cite{non_NLC_ContT}, but with a pre-orientation $\Psi_0\approx 30°$ with respect to $z$ in the $(y,z)$ plane to make a voltage bias unnecessary \cite{lc_ca_alber}. We carried out the experiments with a near infrared ($\lambda=1.064\mu$m) source and high resolution CCD cameras for imaging both the soliton output profiles and their propagation in $(y,z)$.  At the input, two extraordinary-wave solitons were excited with opposite momenta along $x$, compensating for the walk-off in $(y, z)$. Particular attention was paid to the tilt in $(x,z)$ in order to achieve spiraling with (invariant) separation $d$ between the two ``particles''.\\
The experimental results are summarized in Fig. \ref{spir0}, showing images of light propagating in $(y,z)$ (Fig. \ref{spir0} a-d) and output intensity profiles in $(x,y)$ (Fig. \ref{spir0} e-h) for increasing excitation  $P_\mathrm{in}$. In agreement with our theoretical prediction, as the power increased from $2.1$ (Fig. \ref{spir0}a,e) to $3.9mW$ (Fig. \ref{spir0}d-h), the cluster AM changed as well, as witnessed by the rigid rotation of $\approx 180$ degrees in the output plane (Fig. \ref{spir0}e-h). Remarkably enough, each soliton profile remained nearly unmodified (Fig. \ref{spir1}a) and the rotation angle $\delta$ evolved linearly with power (Fig. \ref{spir1}b), demonstrating the nonlinear control over the overall AM angular momentum and in excellent agreement with the model. Owing to the giant reorientational nonlinearity of NLC, the resulting sensitivity $\Delta\delta/\Delta P_\mathrm{in}=\pi/2$ (rad)/mW is quite substantial.

\paragraph{Conclusions. ---}
Stemming from first principles and without specific assumptions on the dielectric response, we theoretically disclosed and experimentally demonstrated an original approach to gain nonlinear control over the angular momentum of a cluster of (two) spatial solitons. Such nonlinear management of the soliton-interaction potential is a remarkable example of all-optical control over light-induced guided-wave interconnect. Due to the general character of the model (and the experimental results) we derived, our findings are going to affect other areas where solitons are actively investigated, including plasma physics and Bose-Einstein condensates, including novel  applications in nonlinear optics. 
\paragraph{Acknowledgement. ---}
The authors thank M. Kaczmarek (U. Southampton) and C. Umeton (U. Calabria). This work was funded in part by the Italian Ministry for University and Research (PRIN 2005098337).


\begin{thebibliography}{48}
\expandafter\ifx\csname natexlab\endcsname\relax\def\natexlab#1{#1}\fi
\expandafter\ifx\csname bibnamefont\endcsname\relax
  \def\bibnamefont#1{#1}\fi
\expandafter\ifx\csname bibfnamefont\endcsname\relax
  \def\bibfnamefont#1{#1}\fi
\expandafter\ifx\csname citenamefont\endcsname\relax
  \def\citenamefont#1{#1}\fi
\expandafter\ifx\csname url\endcsname\relax
  \def\url#1{\texttt{#1}}\fi
\expandafter\ifx\csname urlprefix\endcsname\relax\def\urlprefix{URL }\fi
\providecommand{\bibinfo}[2]{#2}
\providecommand{\eprint}[2][]{\url{#2}}

\bibitem[{\citenamefont{Block et~al.}(2004)\citenamefont{Block, Puerari,
  Freeman, Groess, and Block}}]{PBTMOCD}
\bibinfo{author}{\bibfnamefont{D.~L.} \bibnamefont{Block}},
  \bibinfo{author}{\bibfnamefont{I.}~\bibnamefont{Puerari}},
  \bibinfo{author}{\bibfnamefont{K.~C.} \bibnamefont{Freeman}},
  \bibinfo{author}{\bibfnamefont{R.}~\bibnamefont{Groess}}, \bibnamefont{and}
  \bibinfo{author}{\bibfnamefont{E.~K.} \bibnamefont{Block}},
  \emph{\bibinfo{title}{Penetrating Bars Through Masks of Cosmic Dust: The
  Hubble Tuning Fork Strikes a New Note}} (\bibinfo{publisher}{Springer},
  \bibinfo{address}{Dordrecht}, \bibinfo{year}{2004});
\bibinfo{author}{\bibfnamefont{W.~K.} \bibnamefont{Hartmann}} \bibnamefont{and}
  \bibinfo{author}{\bibfnamefont{S.~M.} \bibnamefont{Larson}},
  \bibinfo{journal}{Icarus} \textbf{\bibinfo{volume}{7}}, \bibinfo{pages}{257}
  (\bibinfo{year}{1967});
\bibinfo{author}{\bibfnamefont{F.~F.} \bibnamefont{Fish}},
  \bibinfo{journal}{Icarus} \textbf{\bibinfo{volume}{7}}, \bibinfo{pages}{251}
  (\bibinfo{year}{1967}).

\bibitem[{\citenamefont{Kazanas}(1977)}]{am_bf_kazana}
\bibinfo{author}{\bibfnamefont{D.}~\bibnamefont{Kazanas}},
  \bibinfo{journal}{Nature} \textbf{\bibinfo{volume}{267}},
  \bibinfo{pages}{501} (\bibinfo{year}{1977});
\bibinfo{author}{\bibfnamefont{I.}~\bibnamefont{Ferrin}},
  \bibinfo{journal}{Nature} \textbf{\bibinfo{volume}{333}},
  \bibinfo{pages}{834} (\bibinfo{year}{1988});
\bibinfo{author}{\bibfnamefont{C.~K.} \bibnamefont{Goertz}} {\bibnamefont{\emph{et~al.}}}, \bibinfo{journal}{Nature}
  \textbf{\bibinfo{volume}{320}}, \bibinfo{pages}{141} (\bibinfo{year}{1986});
\bibinfo{author}{\bibfnamefont{H.~P.} \bibnamefont{Jakobsen}},
  \bibinfo{author}{\bibfnamefont{M.}~\bibnamefont{Kon}}, \bibnamefont{and}
  \bibinfo{author}{\bibfnamefont{I.~E.} \bibnamefont{Segal}},
  \bibinfo{journal}{Phys. Rev. Lett.} \textbf{\bibinfo{volume}{42}},
  \bibinfo{pages}{1788} (\bibinfo{year}{1979}).

\bibitem[{\citenamefont{Mcewan}(1976)}]{am_bf_mcewaa}
\bibinfo{author}{\bibfnamefont{A.~D.} \bibnamefont{Mcewan}},
  \bibinfo{journal}{Nature} \textbf{\bibinfo{volume}{260}},
  \bibinfo{pages}{126} (\bibinfo{year}{1976}).

\bibitem[{\citenamefont{Hide and et~al.}(1980)}]{bf_am_hide}
\bibinfo{author}{\bibfnamefont{R.}~\bibnamefont{Hide}} {\bibnamefont{\emph{et~al.}}}, \bibinfo{journal}{Nature}
  \textbf{\bibinfo{volume}{286}}, \bibinfo{pages}{114} (\bibinfo{year}{1980});
\bibinfo{author}{\bibfnamefont{R.~B.} \bibnamefont{Langley}} {\bibnamefont{\emph{et~al.}}}, \bibinfo{journal}{Nature}
  \textbf{\bibinfo{volume}{294}}, \bibinfo{pages}{730} (\bibinfo{year}{1981}).

\bibitem[{\citenamefont{van~den Oord and et~al.}(1987)}]{bf_am_oord}
\bibinfo{author}{\bibnamefont{van~den Oord}} {\bibnamefont{\emph{et~al.}}}, \bibinfo{journal}{Phys. Rev. Lett.}
  \textbf{\bibinfo{volume}{59}}, \bibinfo{pages}{2907} (\bibinfo{year}{1987}).

\bibitem[{\citenamefont{Torner}(2002)}]{Torner_PRA_03}
\bibinfo{author}{\bibfnamefont{J.~P.} \bibnamefont{Torres}}   \bibinfo{author}{\bibfnamefont{A.}~\bibnamefont{Alexandrescu}} \bibnamefont{and}  \bibinfo{author}{\bibfnamefont{L.}~\bibnamefont{Torner}}, \bibinfo{journal}{Phys. Rev. A}
  \textbf{\bibinfo{volume}{68}}, \bibinfo{pages}{050301(R)} (\bibinfo{year}{2003}).

\bibitem[{\citenamefont{Tabosa and Petrov}(1999)}]{am_mmc_tabosa}
\bibinfo{author}{\bibfnamefont{J.~W.~R.} \bibnamefont{Tabosa}}
  \bibnamefont{and} \bibinfo{author}{\bibfnamefont{D.~V.}
  \bibnamefont{Petrov}}, \bibinfo{journal}{Phys. Rev. Lett.}
  \textbf{\bibinfo{volume}{83}}, \bibinfo{pages}{4967} (\bibinfo{year}{1999});
\bibinfo{author}{\bibfnamefont{N.~B.} \bibnamefont{Simpson}} {\bibnamefont{\emph{et~al.}}}, \bibinfo{journal}{Opt. Lett.}
  \textbf{\bibinfo{volume}{22}}, \bibinfo{pages}{52} (\bibinfo{year}{1997});
\bibinfo{author}{\bibfnamefont{P.}~\bibnamefont{Galajda}} \bibnamefont{and}
  \bibinfo{author}{\bibfnamefont{P.}~\bibnamefont{Ormos}},
  \bibinfo{journal}{Appl. Phys. Lett.} \textbf{\bibinfo{volume}{78}},
  \bibinfo{pages}{249} (\bibinfo{year}{2001});
\bibinfo{author}{\bibfnamefont{E.}~\bibnamefont{Santamato}} {\bibnamefont{\emph{et~al.}}}, \bibinfo{journal}{Phys. Rev. Lett.}
  \textbf{\bibinfo{volume}{57}}, \bibinfo{pages}{2423} (\bibinfo{year}{1986});
\bibinfo{author}{\bibfnamefont{T.~V.} \bibnamefont{Galstyan}} \bibnamefont{and}
  \bibinfo{author}{\bibfnamefont{V.}~\bibnamefont{Drnoyan}},
  \bibinfo{journal}{Phys. Rev. Lett.} \textbf{\bibinfo{volume}{78}},
  \bibinfo{pages}{2760} (\bibinfo{year}{1997}).

\bibitem[{\citenamefont{Minns and et~al.}(2003)}]{bf_amc_minns}
\bibinfo{author}{\bibfnamefont{R.~S.} \bibnamefont{Minns}} {\bibnamefont{\emph{et~al.}}}, \bibinfo{journal}{Phys. Rev. Lett.}
  \textbf{\bibinfo{volume}{91}}, \bibinfo{eid}{243601} (\bibinfo{year}{2003});
\bibinfo{author}{\bibfnamefont{N.}~\bibnamefont{Dudovich}},
  \bibinfo{author}{\bibfnamefont{D.}~\bibnamefont{Oron}}, \bibnamefont{and}
  \bibinfo{author}{\bibfnamefont{Y.}~\bibnamefont{Silberberg}},
  \bibinfo{journal}{Phys. Rev. Lett.} \textbf{\bibinfo{volume}{92}},
  \bibinfo{eid}{103003} (\bibinfo{year}{2004}).

\bibitem[{\citenamefont{Mair et~al.}(2001)\citenamefont{Mair, Vaziri, Weihs,
  and Zeilinger}}]{bf_am_maira}
\bibinfo{author}{\bibfnamefont{A.}~\bibnamefont{Mair}}
{\bibnamefont{\emph{et~al.}}},
  \bibinfo{journal}{Nature} \textbf{\bibinfo{volume}{412}},
  \bibinfo{pages}{313} (\bibinfo{year}{2001});
\bibinfo{author}{\bibfnamefont{G.}~\bibnamefont{Molina-Terriza}},
  \bibinfo{author}{\bibfnamefont{J.~P.} \bibnamefont{Torres}},
  \bibnamefont{and} \bibinfo{author}{\bibfnamefont{L.}~\bibnamefont{Torner}},
  \bibinfo{journal}{Phys. Rev. Lett.} \textbf{\bibinfo{volume}{88}},
  \bibinfo{pages}{013601} (\bibinfo{year}{2001}).

\bibitem[{\citenamefont{Whitham}(1999)}]{LANW}
\bibinfo{author}{\bibfnamefont{G.~B.} \bibnamefont{Whitham}},
  \emph{\bibinfo{title}{Linear and Nonlinear Waves}}
  (\bibinfo{publisher}{Wiley}, \bibinfo{address}{New York},
  \bibinfo{year}{1999}).

\bibitem[{\citenamefont{Fermi et~al.}(1955, Unpublished)\citenamefont{Fermi,
  Pasta, and Ulam}}]{sol_bf_fermia}
\bibinfo{author}{\bibfnamefont{E.}~\bibnamefont{Fermi}},
  \bibinfo{author}{\bibfnamefont{J.}~\bibnamefont{Pasta}}, \bibnamefont{and}
  \bibinfo{author}{\bibfnamefont{S.}~\bibnamefont{Ulam}}, \bibinfo{journal}{Los
  Alamos Report} \textbf{\bibinfo{volume}{LA-194}} (\bibinfo{year}{1955,
  Unpublished});
\bibinfo{author}{\bibfnamefont{E.}~\bibnamefont{Segre}},
  \emph{\bibinfo{title}{Collected Papers of Enrico Fermi}}
  (\bibinfo{publisher}{University of Chicago Press},
  \bibinfo{address}{Chicago}, \bibinfo{year}{1965}).

\bibitem[{\citenamefont{Bishop and Schneider}(1978)}]{SICMP}
\bibinfo{author}{\bibfnamefont{A.}~\bibnamefont{Bishop}} \bibnamefont{and}
  \bibinfo{author}{\bibfnamefont{T.}~\bibnamefont{Schneider}},
  \emph{\bibinfo{title}{Solitons and Condensed Matter Physics}}
  (\bibinfo{publisher}{Springer}, \bibinfo{address}{Berlin},
  \bibinfo{year}{1978}).

\bibitem[{\citenamefont{Zabusky and Kruskal}(1965)}]{sol_pl_Zabus}
\bibinfo{author}{\bibfnamefont{N.~J.} \bibnamefont{Zabusky}} \bibnamefont{and}
  \bibinfo{author}{\bibfnamefont{M.~D.} \bibnamefont{Kruskal}},
  \bibinfo{journal}{Phys. Rev. Lett.} \textbf{\bibinfo{volume}{15}},
  \bibinfo{pages}{240} (\bibinfo{year}{1965}).

\bibitem[{\citenamefont{Burger and et~al.}(1999)}]{sol_bec_Burge}
\bibinfo{author}{\bibfnamefont{S.}~\bibnamefont{Burger}} {\bibnamefont{\emph{et~al.}}}, \bibinfo{journal}{Phys. Rev. Lett.}
  \textbf{\bibinfo{volume}{83}}, \bibinfo{pages}{5198} (\bibinfo{year}{1999});
\bibinfo{author}{\bibfnamefont{J.}~\bibnamefont{Denschlag}} {\bibnamefont{\emph{et~al.}}}, \bibinfo{journal}{Science}
  \textbf{\bibinfo{volume}{287}}, \bibinfo{pages}{97} (\bibinfo{year}{2000}).

\bibitem[{\citenamefont{Kivshar and Agrawal}(2003)}]{OSFFTPC}
\bibinfo{author}{\bibfnamefont{G.I.}~\bibnamefont{Stegeman}} \bibnamefont{and} \bibinfo{author}{\bibfnamefont{M.} \bibnamefont{Segev}}, \bibinfo{journal}{Science}
  \textbf{\bibinfo{volume}{286}}, \bibinfo{pages}{1518} (\bibinfo{year}{1999});
\bibinfo{author}{\bibfnamefont{S.}~\bibnamefont{Trillo}} \bibnamefont{and}
  \bibinfo{author}{\bibfnamefont{W.~E.} \bibnamefont{Torruellas}},
  \emph{\bibinfo{title}{Spatial Solitons}} (\bibinfo{publisher}{Springer},
  \bibinfo{address}{Berlin}, \bibinfo{year}{2001});
\bibinfo{author}{\bibfnamefont{Y.~S.} \bibnamefont{Kivshar}} \bibnamefont{and}
  \bibinfo{author}{\bibfnamefont{G.~P.} \bibnamefont{Agrawal}},
  \emph{\bibinfo{title}{Optical Solitons: from Fibers to Photonic Crystals}}
  (\bibinfo{publisher}{Academic Press}, \bibinfo{address}{San Diego},
  \bibinfo{year}{2003}).

 \bibitem[{\citenamefont{Hoffmann}(2002)}]{Hoffmann}
\bibinfo{author}{\bibfnamefont{R.}~\bibnamefont{Hoffmann}}
 \emph{\bibinfo{title}{Soliton}}
  (\bibinfo{publisher}{Truman State University Press}, \bibinfo{address}{Kirksville},
  \bibinfo{year}{2002}).
  
\bibitem[{\citenamefont{Belinski and Verdaguer}(2001)}]{GS}
\bibinfo{author}{\bibfnamefont{V.}~\bibnamefont{Belinski}} \bibnamefont{and}
  \bibinfo{author}{\bibfnamefont{E.}~\bibnamefont{Verdaguer}},
  \emph{\bibinfo{title}{Gravitational Solitons}} (\bibinfo{publisher}{Cambridge
  Press}, \bibinfo{address}{Cambridge}, \bibinfo{year}{2001}).

\bibitem[{\citenamefont{Toda}(1983)}]{NWAS}
\bibinfo{author}{\bibfnamefont{M.}~\bibnamefont{Toda}},
  \emph{\bibinfo{title}{Nonlinear Waves and Solitons}}
  (\bibinfo{publisher}{KTK}, \bibinfo{address}{Tokio}, \bibinfo{year}{1983});
\bibinfo{author}{\bibfnamefont{A.~C.} \bibnamefont{Newell}},
  \emph{\bibinfo{title}{Solitons in Mathematics and Physics}}
  (\bibinfo{publisher}{SIAM}, \bibinfo{address}{Arizona},
  \bibinfo{year}{1985}).

\bibitem[{\citenamefont{Bigelow et~al.}(2004)\citenamefont{Bigelow, Zerom, and
  Boyd}}]{fl_bf_bigela}
\bibinfo{author}{\bibfnamefont{M.~S.} \bibnamefont{Bigelow}},
  \bibinfo{author}{\bibfnamefont{P.}~\bibnamefont{Zerom}}, \bibnamefont{and}
  \bibinfo{author}{\bibfnamefont{R.~W.} \bibnamefont{Boyd}},
  \bibinfo{journal}{Phys. Rev. Lett.} \textbf{\bibinfo{volume}{92}},
  \bibinfo{eid}{083902} (\bibinfo{year}{2004});
\bibinfo{author}{\bibfnamefont{W.~J.} \bibnamefont{Firth}} \bibnamefont{and}
  \bibinfo{author}{\bibfnamefont{D.~V.} \bibnamefont{Skryabin}},
  \bibinfo{journal}{Phys. Rev. Lett.} \textbf{\bibinfo{volume}{79}},
  \bibinfo{pages}{2450} (\bibinfo{year}{1997}).
  
\bibitem[{\citenamefont{Mihalache and et~al.}(2002)}]{Mih_2002}
\bibinfo{author}{\bibfnamefont{D.}~\bibnamefont{Mihalache}} {\bibnamefont{\emph{et~al.}}}, \bibinfo{journal}{Phys. Rev. Lett.} \textbf{\bibinfo{volume}{88}}, \bibinfo{pages}{073902} (\bibinfo{year}{2002}).

\bibitem[{\citenamefont{Beli\ifmmode~\acute{c}\else \'{c}\fi{}
  et~al.}(1999)\citenamefont{Beli\ifmmode~\acute{c}\else \'{c}\fi{}, Stepken,
  and Kaiser}}]{sp_bf_belica}
\bibinfo{author}{\bibfnamefont{M.~R.} \bibnamefont{Beli\ifmmode~\acute{c}\else
  \'{c}\fi{}}}, \bibinfo{author}{\bibfnamefont{A.}~\bibnamefont{Stepken}},
  \bibnamefont{and} \bibinfo{author}{\bibfnamefont{F.}~\bibnamefont{Kaiser}},
  \bibinfo{journal}{Phys. Rev. Lett.} \textbf{\bibinfo{volume}{82}},
  \bibinfo{pages}{544} (\bibinfo{year}{1999});
\bibinfo{author}{\bibfnamefont{A.~V.} \bibnamefont{Buryak}} {\bibnamefont{\emph{et~al.}}}, \bibinfo{journal}{Phys. Rev. Lett.}
  \textbf{\bibinfo{volume}{82}}, \bibinfo{pages}{81} (\bibinfo{year}{1999});
\bibinfo{author}{\bibfnamefont{M.-F.} \bibnamefont{Shih}},
  \bibinfo{author}{\bibfnamefont{M.}~\bibnamefont{Segev}}, \bibnamefont{and}
  \bibinfo{author}{\bibfnamefont{G.}~\bibnamefont{Salamo}},
  \bibinfo{journal}{Phys. Rev. Lett.} \textbf{\bibinfo{volume}{78}},
  \bibinfo{pages}{2551} (\bibinfo{year}{1997}).

\bibitem[{\citenamefont{Rotschild and et~al.}(2006)}]{to_ss_rotsca}
\bibinfo{author}{\bibfnamefont{C.} \bibnamefont{Rotschild}} {\bibnamefont{\emph{et~al.}}}, \bibinfo{journal}{Nature Physics}
  \textbf{\bibinfo{volume}{2}}, \bibinfo{pages}{769} (\bibinfo{year}{2006});

\bibitem[{\citenamefont{Desyatnikov and Kivshar}(2002)}]{bf_sc_desyaa}
\bibinfo{author}{\bibfnamefont{A.~S.} \bibnamefont{Desyatnikov}}
  \bibnamefont{and} \bibinfo{author}{\bibfnamefont{Y.~S.}
  \bibnamefont{Kivshar}}, \bibinfo{journal}{Phys. Rev. Lett.}
  \textbf{\bibinfo{volume}{88}}, \bibinfo{pages}{053901}
  (\bibinfo{year}{2002}).

\bibitem[{\citenamefont{Olver}(1986)}]{AOLGTDE}
\bibinfo{author}{\bibfnamefont{P.~J.} \bibnamefont{Olver}},
  \emph{\bibinfo{title}{Applications of Lie groups to differential equations}}
  (\bibinfo{publisher}{Springer-Verlag}, \bibinfo{address}{New York},
  \bibinfo{year}{1986}).

\bibitem[{\citenamefont{Peccianti and et~al.}(2004)}]{nlc_st_pecci}
\bibinfo{author}{\bibfnamefont{M.}~\bibnamefont{Peccianti}} {\bibnamefont{\emph{et~al.}}}, \bibinfo{journal}{Appl. Phys. Lett.} \textbf{\bibinfo{volume}{77}}, \bibinfo{pages}{7} (\bibinfo{year}{2000});
\bibinfo{author}{\bibfnamefont{M.}~\bibnamefont{Peccianti}} {\bibnamefont{\emph{et~al.}}}, \bibinfo{journal}{Nature}
  \textbf{\bibinfo{volume}{432}}, \bibinfo{pages}{733} (\bibinfo{year}{2004});
\bibinfo{author}{\bibfnamefont{M.}~\bibnamefont{Peccianti}} {\bibnamefont{\emph{et~al.}}}, \bibinfo{journal}{Nature Physics}  \textbf{\bibinfo{volume}{2}}, \bibinfo{pages}{737} (\bibinfo{year}{2006}).
  
\bibitem[{\citenamefont{Bang and et~al.}(2002)}]{non_BangS}
\bibinfo{author}{\bibfnamefont{O.}~\bibnamefont{Bang}} {\bibnamefont{\emph{et~al.}}}, \bibinfo{journal}{Phys. Rev. E}
  \textbf{\bibinfo{volume}{66}}, \bibinfo{pages}{046619}
  (\bibinfo{year}{2002}).

\bibitem[{\citenamefont{Conti et~al.}(2003)\citenamefont{Conti, Peccianti, and
  Assanto}}]{non_NLC_ContT}
\bibinfo{author}{\bibfnamefont{C.}~\bibnamefont{Conti}},
  \bibinfo{author}{\bibfnamefont{M.}~\bibnamefont{Peccianti}},
  \bibnamefont{and} \bibinfo{author}{\bibfnamefont{G.}~\bibnamefont{Assanto}},
  \bibinfo{journal}{Phys. Rev. Lett.} \textbf{\bibinfo{volume}{91}},
  \bibinfo{pages}{073901} (\bibinfo{year}{2003}); ibid. \textbf{\bibinfo{volume}{92}},
  \bibinfo{pages}{113902} (\bibinfo{year}{2004}).

\bibitem[{\citenamefont{Zakharov}(1990)}]{WII}
\bibinfo{author}{\bibfnamefont{V.~E.} \bibnamefont{Zakharov}},
  \emph{\bibinfo{title}{What is integrability?}}
  (\bibinfo{publisher}{Springer}, \bibinfo{address}{Heidelberg},
  \bibinfo{year}{1990}).

\bibitem[{\citenamefont{Christodoulides and Carvalho}(1994)}]{pr_nl_christ}
\bibinfo{author}{\bibfnamefont{D.~N.} \bibnamefont{Christodoulides}}
  \bibnamefont{and} \bibinfo{author}{\bibfnamefont{M.~I.}
  \bibnamefont{Carvalho}}, \bibinfo{journal}{Opt. Lett.}
  \textbf{\bibinfo{volume}{19}}, \bibinfo{pages}{1714} (\bibinfo{year}{1994}).

\bibitem[{\citenamefont{Conti}(2005)}]{bf_cpxl_conti}
\bibinfo{author}{\bibfnamefont{C.}~\bibnamefont{Conti}},
  \bibinfo{journal}{Phys. Rev. E} \textbf{\bibinfo{volume}{72}},
  \bibinfo{eid}{066620} (\bibinfo{year}{2005}).

\bibitem[{\citenamefont{Conti}(2005)}]{pecciante_OL}
\bibinfo{author}{\bibfnamefont{C.}~\bibnamefont{Conti}}, 
\bibinfo{author}{\bibfnamefont{M.}
  \bibnamefont{Peccianti}} \bibnamefont{and} \bibinfo{author}{\bibfnamefont{G.}
  \bibnamefont{Assanto}},
  \bibinfo{journal}{Opt. Lett.} \textbf{\bibinfo{volume}{31}},
  \bibinfo{eid}{2030} (\bibinfo{year}{2006}).

\bibitem[{\citenamefont{Perez-Garcia and Vekslerchik}(2003)}]{bf_sc_garcia}
\bibinfo{author}{\bibfnamefont{V.~M.} \bibnamefont{Perez-Garcia}}
  \bibnamefont{and}
  \bibinfo{author}{\bibfnamefont{V.}~\bibnamefont{Vekslerchik}},
  \bibinfo{journal}{Phys. Rev. E} \textbf{\bibinfo{volume}{67}},
  \bibinfo{eid}{061804} (\bibinfo{year}{2003}).

\bibitem[{\citenamefont{El and Kamchatnov}(2005)}]{bf_solg_ela}
\bibinfo{author}{\bibfnamefont{G.~A.} \bibnamefont{El}} \bibnamefont{and}
  \bibinfo{author}{\bibfnamefont{A.~M.} \bibnamefont{Kamchatnov}},
  \bibinfo{journal}{Phys. Rev. Lett.} \textbf{\bibinfo{volume}{95}},
  \bibinfo{pages}{204101} (\bibinfo{year}{2005}).

\bibitem[{\citenamefont{Alberucci and et~al.}(2005)}]{lc_ca_alber}
\bibinfo{author}{\bibfnamefont{A.}~\bibnamefont{Alberucci}} \bibinfo{author}{\bibnamefont{\emph{et~al.}}}, \bibinfo{journal}{Opt. Lett.}
  \textbf{\bibinfo{volume}{30}}, \bibinfo{pages}{1381} (\bibinfo{year}{2005}).
  

\end{thebibliography}

\end{document}